\documentclass[manuscript,screen]{acmart}

\AtBeginDocument{%
  \providecommand\BibTeX{{%
    \normalfont B\kern-0.5em{\scshape i\kern-0.25em b}\kern-0.8em\TeX}}}

\setcopyright{acmcopyright}
\copyrightyear{2024}
\acmYear{2024}
\acmDOI{XXXXXXX.XXXXXXX}

\acmConference[CHI '23 GenAICHI Workshop]{Make}{Apr. 2023}{Yongquan et al.}
\acmBooktitle{Proc. ACM Conference on Human Factors in Computing Systems (CHI): Generative AI and HCI Workshop 2023} 
\acmPrice{15.00}
\acmISBN{978-1-4503-XXXX-X/18/06}

\usepackage{multirow}
\usepackage{graphicx}
\usepackage{subfigure} 
\usepackage{xcolor,soul}
\usepackage{array}
\usepackage{booktabs}
\usepackage{tabularx}

\begin{document}


\title[Investigating the Design Considerations for Integrating Text-to-Image Generative AI within Augmented Reality Environments]{Investigating the Design Considerations for Integrating Text-to-Image Generative AI within Augmented Reality Environments}

\author{Yongquan Hu}
 \email{yongquan.hu@unsw.edu.au}
\affiliation{%
 \institution{University of New South Wales}
 \country{Australia}}

 \author{Dawen Zhang}
\email{david.zhang@data61.csiro.au}
\affiliation{%
 \institution{CSIRO's Data61}
 \country{Australia}}

  \author{Mingyue Yuan}
\email{mingyue.yuan@unsw.edu.au}
\affiliation{%
 \institution{University of New South Wales}
 \country{Australia}}

  \author{Kaiqi Xian}
\email{kaiqi.xian@student.unsw.edu.au}
\affiliation{%
 \institution{University of New South Wales}
 \country{Australia}}

   \author{Don Samitha Elvitigala}
\email{don.elvitigala@monash.edu}
\affiliation{%
 \institution{Monash University}
 \country{Australia}}

\author{June Kim}
\email{june.kim@unsw.edu.au}
\affiliation{%
 \institution{University of New South Wales}
 \country{Australia}}

\author{Gelareh Mohammadi}
\email{g.mohammadi@unsw.edu.au}
\affiliation{%
 \institution{University of New South Wales}
 \country{Australia}}

 \author{Zhenchang Xing}
\email{Zhenchang.Xing@data61.csiro.au}
\affiliation{%
 \institution{CSIRO's Data61}
 \country{Australia}}

 \author{Xiwei Xu}
\email{Xiwei.Xu@data61.csiro.au}
\affiliation{%
 \institution{CSIRO's Data61}
 \country{Australia}}

 \author{Aaron Quigley}
\email{aquigley@acm.org}
\affiliation{%
 \institution{CSIRO's Data61}
 \country{Australia}}

\renewcommand{\shortauthors}{Yongquan et al.}

\begin{abstract}


Generative Artificial Intelligence (GenAI) has emerged as a fundamental component of intelligent interactive systems, enabling the automatic generation of multimodal media content. The continuous enhancement in the quality of Artificial Intelligence-Generated Content (AIGC), including but not limited to images and text, is forging new paradigms for its application, particularly within the domain of Augmented Reality (AR). Nevertheless, the application of GenAI within the AR design process remains opaque. This paper aims to articulate a design space encapsulating a series of criteria and a prototypical process to aid practitioners in assessing the aptness of adopting pertinent technologies. The proposed model has been formulated based on a synthesis of design insights garnered from ten experts, obtained through focus group interviews. Leveraging these initial insights, we delineate potential applications of GenAI in AR.

\end{abstract}

\ccsdesc[500]{Human-centered computing~Human computer interaction (HCI)}
\ccsdesc[300]{Empirical Study}

\keywords{Augmented Reality; Prompt Engineering; Generative AI; Content Creation.}

\maketitle

\section{Introduction \& Related Work}

Augmented Reality (AR) serves as a means to connect the physical and digital worlds, supplementing or extending the former with rich decorative and informative visual effects. \cite{jung2018augmented, nebeling2018trouble}. 
There are three major variations for AR depending on the display methods: Spatial Augmented Reality (SAR), Head-Mounted Display (HMD), and Hand-Held Display (HHD) \cite{wang2021ar, giunta2018review}.
Each of them presents particular advantages and disadvantages for various tasks and scenarios, as examined extensively in previous studies. 
For instance, SAR has been identified to merge the real and virtual worlds by directly projecting light onto physical surfaces, but it poses privacy concerns as the displayed content is publicly visible \cite{roesner2014security, kotsios2015privacy}.
Whilst HHD and HMD offer considerable advantages in terms of privacy protection, they inadvertently engender a somewhat isolated user experience that may impede the sharing of content \cite{birlo2022utility}.
Researchers have also explored the integration of multiple display technologies.
For instance, Hartmann et al. proposed Augmented Augmented Reality (AAR) by combining wearable AR displays with wearable spatial augmented reality projectors to mitigate the isolated experience of using an individual AR device \cite{hartmann2020aar}.
All of the different methods are meant to offer enhanced contextual information and immersion while retaining focus on the physical world for users.


Two major components in AR content are images and text \cite{bach2017drawing, weerasinghe2022arigato, chiu2018interactive, jing2019snapchart}, both of which currently require human involvement in their creation, such as 3D modeling using Unity or animation story scripting \cite{bassyouni2021augmented, vaataja2013exploring}.
However, recent advancements in Artificial Intelligence (AI) technology have improved the capabilities of AI-generated content (AIGC) to the extent that the boundary between human and machine-generated content has become impressively indistinguishable. 
For instance, the Large Language Models (LLM), such as GPT and its variants \cite{chatgpt} and PaLM \cite{chowdhery2022palm}, are capable of generating high-quality conversational responses or completing text contextually\cite{dale2021gpt, floridi2020gpt}.
Another type of recently prevalent generative model is text-to-image models, such as the Stable Diffusion \cite{rombach2022high}, Disco Diffusion \cite{rafael2021diffusion} and DALL·E 2 \cite{marcus2022very}, which generate artistic images by given textual prompts.
These generative models have drawn major attention from various academic and industrial fields.
Diverse attempts are being conducted to utilize them to facilitate the daily routine and working process (e.g., code generation, documentation translation, and illustration generation), but the study of the combination of generative AI and AR (AIGC+AR) remains inactive and remains room for discovery.
There are some existing works that applied generative AI models to specific AR scenarios, focusing on technical solutions or engineering tasks. For example, Asangika et al. proposed ARGAN \cite{sandamini2022augmented}, an Augmented Reality-based Fashion Design system, leverages Controllable Generative Adversarial Networks to generate new attire from a sketch and theme image, allowing real-time visualization of virtual 2D apparel on a human body.
However, the discussion on the generic design guidance and holistic system analysis of generative AI in AR has not been systemically stated yet\cite{xu2023generative}, especially concerning the general discussion of design space where AIGC is employed in AR display.
Such guidance and discussion are crucial for the design decision-making whenever the system involves the usage of AIGC in AR, because the designs would dramatically impact the user experience and determine the success of the application.
Therefore, in this paper, we will introduce a design space that concerns multiple factors in the design progress.
This design space is drawn from an in-depth focus group interview involving 10 experts who are instructed to use our AIGC+AR prototype and answer a set of questions.
In particular, our contributions are as follows: 

    \begin{itemize}
    \item A prototype applying two prevalent LLM and text-to-image models in three AR variations to allow the user to experience different AIGC+AR designs.
    \item Focus groups and discussion summary on a ``user-function-environment'' design thinking.
    \item Potential application scenarios for the combination of AIGC and AR.
    \end{itemize}

\begin{figure*}
\centering
\includegraphics[width=0.9\textwidth]{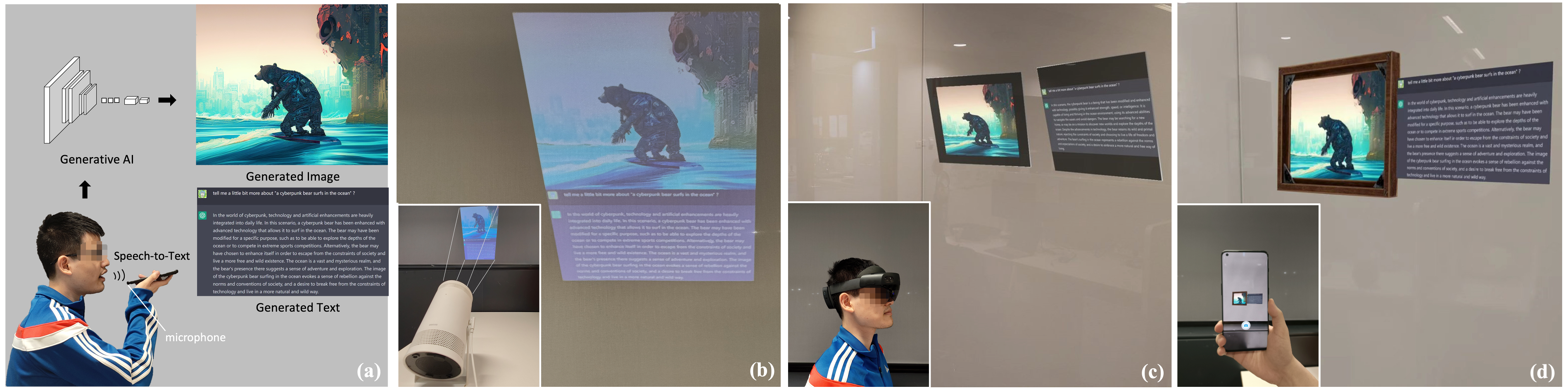}
  \caption{The workflow and display effect of our GenerativeAIR prototype: (a) The user's speech into the microphone is converted into text, which is then fed into an AI model for generating artistic images and more text; (b) Generated content in Spatial Augmented Reality (SAR): an example of Samsung Freestyle project; (c) Generated content in Head-Mounted Display (HMD): an example of Microsoft HoloLens 2; (d) Generated content in Hand-Held Display (HHD): an example of OnePlus 10 Pro.}
  ~\label{fig:prototype}
\end{figure*}

\section{Prototype Implementation}\label{prototype}
In order to enhance the more intuitive experience of subsequent interviewees and obtain more design references, we have developed an experimental prototype system called ``GenerativeAIR'' (Generative AI plus AR). It could be seen as an instance of AIGC+AR to be explored.
The system comprises of diverse generative models (two multimodal generative AI models) and AR devices (three AR display devices).
It takes speech as the interactive input of the system and then generates text and image contents through the generative models that will be displayed on different AR devices.
Regarding the specific AR devices, we use Samsung Freestyle projector \cite{quin2022reviews}  for SAR, HoloLens 2 \cite{hololens2} for HMD, and Oneplus 10 Pro \cite{oneplus} for HHD.


In detail, GenerativeAIR first uses the built-in microphone in the mobile phone to convert the voice of the user into text.
Next, it leverages the application programming interface (API) provided by Google is for speech-to-text \cite{speech2text}. 
As for the AI generation part, GenerativeAIR uses ChatGPT \cite{chatgpt} for text-text generation and Stable Diffusion 2 \cite{stablediffusion2} for text-image generation.
Note that these models are mounted on the cloud rather than deployed locally on the phone. 
The prompts for content generation and the generated results (image and text) are both transmitted through the wireless network, which inevitably leads to a certain but tolerable delay (the actual test delay in our network environment is at the millisecond  level).
Figure~\ref{fig:prototype} shows the overall workflow of the system.
In the beginning, the user speaks through a microphone, and then the transcribed text is used as input into two AI models to generate corresponding images and text (as shown in Figure~\ref{fig:prototype} (a)).
Consequently, the generated media content is transmitted to different AR devices through the network and displayed (as shown from Figure~\ref{fig:prototype} (b) to (d)).


\section{Methodology}\label{methodology}
In order to elicit design factors for AIGC+AR in an open-ended fashion, hence, we held an internal interview and brainstormed.
We use focus group interviews in this process, which is particularly suitable for early exploration in identifying new problems and assessing users' needs \cite{morgan1997focus}. 
The participants are allowed to freely use the GenerativeAIR and will be asked a set of questions regarding their experience with the AIGC+AR application.

\subsection{Participants}
We carried out in-depth focus group discussions with a panel of our ten authors, comprised of six males and four females. Participants were identified by the indices P1 to P10, and their backgrounds varied: 6 participants were academic researchers from different disciplines (computer science (4), mechanical engineering (1), and design (1)); and the remaining 4 were working professionals from various industries ( UI/UX design (2), telecommunications (1), and IT (1)). The mean age of the participants was 28.7 years (SD=6.90), and all of them had at least two years of experience studying the technology or design of AI or AR.

\subsection{Procedure}
We conducted three focus groups with a total of ten participants (G1=3; G2=3; G3=4), each lasting approximately 80 minutes and consisting of five steps. Firstly, the moderator introduced the research purpose (\textasciitilde5 mins). Secondly, participants were asked to freely experience the GenerativeAIR system and respond to any questions raised during the process (\textasciitilde15 minutes). Thirdly, participants provided self-introductions and shared their initial impressions of the GenerativeAIR system (\textasciitilde10 mins). Fourthly, In the main discussion participants freely discussed two topics: RQ1) What are the characteristics need to be considered when comparing AIGC and AR with other related technologies; RQ2) what features should be envisioned when developing the AIGC and AR technology itself (\textasciitilde40 mins). Finally, a summary and debriefing of the discussion was provided (\textasciitilde10 minutes). It is worth mentioning that in the fourth step, for the first question RQ1, we sent a questionnaire to the participants, asking them to score and compare AIGC and AR with their related technologies. More details are in the following section \ref{RQ2}.

\subsection{Analysis}
The discussions in each focus group were recorded and later transcribed and coded using Grounded Theory \cite{strauss1998basics} by our two authors. To ensure the validity of the motivation categorization, efforts were made to minimize the influence of less logical statements that are common in focus groups. Specifically, the moderator encouraged participants to reflect on and verbalize the underlying logical meaning behind their statements. During the coding phase, less logical statements without support from other statements were excluded as evidence. Furthermore, a visualization analysis is also involved for the first question RQ1, in order to better clarify the consideration factors in system design.

\begin{figure*}
\centering
\includegraphics[width=0.65\textwidth]{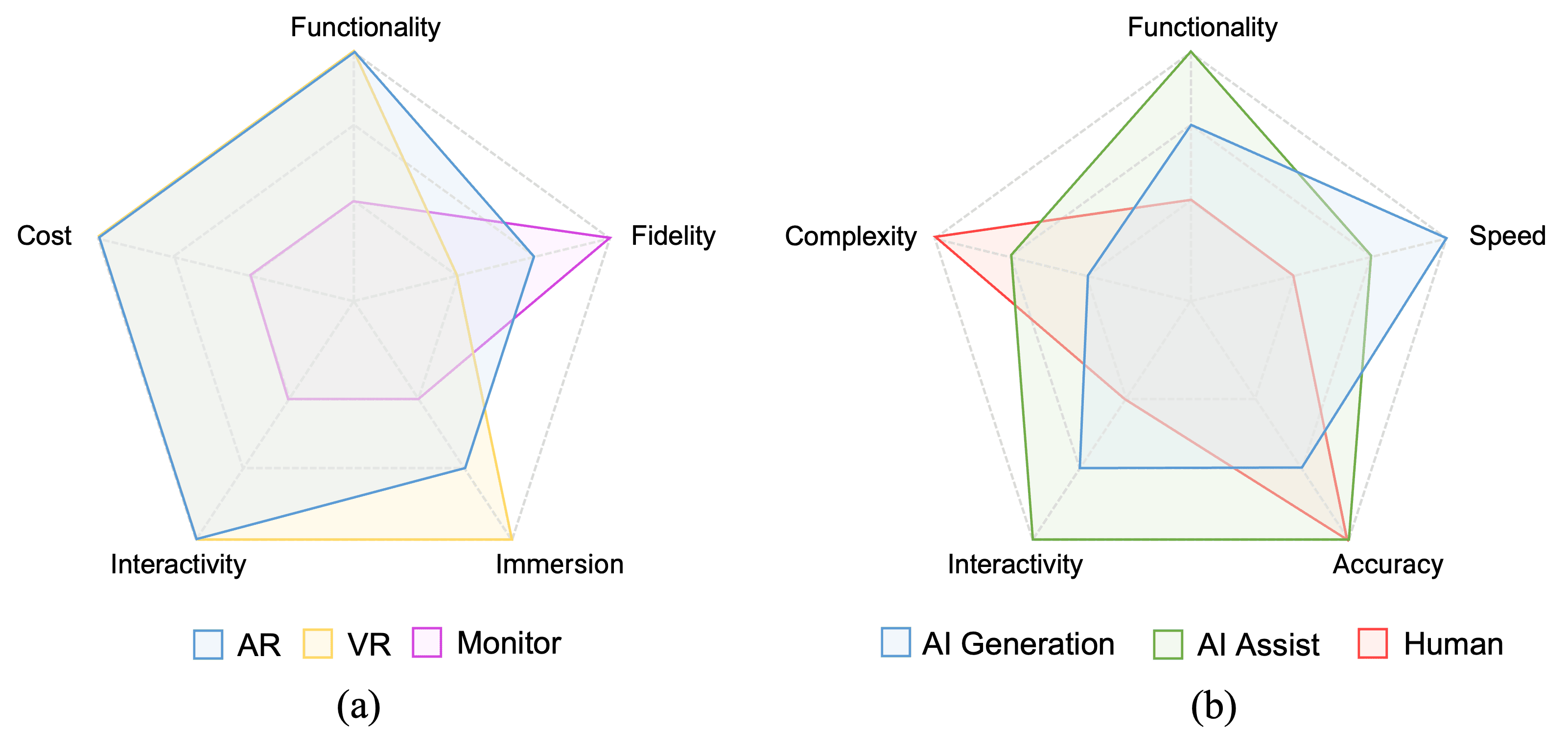}
  \caption{The comparisons of AR display+generative AI and their related techniques: (a) display performance comparison of AR, VR and normal monitor; (b) content-generation performance comparison of generative AI (machine), AI assist (machine+human) and human.}
  ~\label{fig:comparison}
\end{figure*}

\section{Discussions \& Findings}\label{discussion}
In this section, we present and analyze the results of focus group discussions on the design factors that need to be paid attention to when integrating AIGC and AR by answering the following two research questions: the first question is to investigate the importance of the overall system by presenting external characteristics; the second question is for studying the performance of the system itself through clarifying internal differences.

\subsection{RQ1: What are the characteristics need to be considered when comparing AIGC and AR with other related technologies?}\label{RQ1}

In order to identify the advantages and disadvantages of AIGC+AR in comparison to related technologies and discover more suitable application scenarios in the future, a visualization method to qualitatively compare the characteristics of distinct dimension was implemented \cite{feger2022electronicsar}. Given the lack of related work that is similar to the overall system, we separately compared and analyzed its two components, namely, AR display and generative AI. The comparison was based on five different dimensions for both display performance (Figure \ref{fig:comparison} (a)) and content generation performance (Figure \ref{fig:comparison} (b)). The three gray pentagons of different sizes in the figure represent low, medium, and high levels, respectively. It should be noted that these ratings are based on extensive discussions with 10 participants, but they were not exhibited in any systematically reviewed literature. Therefore, these results cannot be considered as rigorous or unique findings, nor are they conclusive regarding the significance of related technologies. Rather, they are intended to address the external characteristics for design considerations.

With respect to display performance, most participants generally believed the AR technology is superior to traditional monitors in terms of functionality, interactivity, and immersion. However, it comes at a higher cost and with lower fidelity. Here, we define that the dimension ``fidelity'' refers to the degree of similarity between the displayed virtual objects and the physical world. For example, participant P2 expressed that using AR displays can obtain more interesting and rich experiences, but the displayed virtual objects are still very different from the physical objects in the real world by saying: \textit{``I was very obsessed with Pokémon GO, a mobile AR game. Its novel operating experience and interesting settings brought me a lot of fun. Yet the display effect of Pokemon in the game is not satisfactory. For example, sometimes Pikachu will appear in the air on the edge of my table or the light and shadow of the displayed trees look strange. It is easy for people to realize that these virtual objects are fake.''}. Considering the current gap between virtual simulation and the real world, some participants thought that a higher proportion of virtual components in the display might reduce the user's real experience of the physical world.  For example, the participant P6 worried that too many 3D virtual objects could aggravate her dizziness and cause discomfort by saying: \textit{``I have 3D motion sickness, so I prefer AR to VR because it's less virtual and I feel better. I'm looking forward to having a custom function for the displayed virtual part, so that I can easily decide its position, size and proportion of the screen.''}. Certainly, there may be more complicated factors in practical cases that need to be considered in-depth in future work.
For content generation performance, all participants agreed that generative AI has an unparalleled advantage over human generation in terms of speed and complexity, whether compared to human generation alone or human-machine collaborative generation. For example, the participant P1 greatly appreciated this convenience for his life by saying \textit{``I am a painting enthusiast. Imagining that you only need to say a few words to the machine to generate a Van Gogh-style Opera House of Sydney. Generative AI is really amazing for me!''}. Participant P2 also believed that this high efficiency improved her work productivity by saying \textit{``Chatgpt can help me program ! I tried to assign some simple code tasks to it, and it can be completed very well, which greatly improved my work efficiency.''}. In addition, we noticed that there was some controversy among the participants regarding the "accuracy" rating in Figure \ref{fig:comparison} (b), as accuracy refers to the gap between the generated content and the expected results of human generation, which is a relatively subjective indicator. Some participants (e.g., P1) felt that AIGC content (such as automatically generated art drawings) was better than self-made ones by saying: \textit{``AI-generated paintings are better than mine''}, while others  (e.g., P5) thought otherwise by saying: \textit{``The layout and storytelling of AI paintings are far from meeting my expectations''}. Hence, we finally hypothesize that, regardless of complicating factors such as time cost or individual ability difference, the most satisfactory results are achieved when people are involved in the generation process. Although the results generated by state-of-the-art AI models are already close to human expectations in some specific cases, we consider the ``medium'' rating to be cautious.

\subsection{RQ2: What features should be envisioned when developing the AIGC and AR technology itself?}\label{RQ2}

The comparison of AR and AIGC and related technologies stimulates ideas for application scenarios. Further exploration of their technical features and details can improve interactive experience and system performance.

Regarding display form, there are 7 participants who all involved remarks with similar meanings: the portability of AR glasses and mobile phones distinguishes them from stationary display methods like SAR. As almost everyone now owns a smartphone, it is expected to be the most common way for following related work on generative AR display. For example, the participant P3  mentioned that mobile phone AR (HHD) has unparalleled advantages in mobility and flexibility compared to the other two AR display methods by saying: \textit{``I'm not always comfortable wearing AR glasses, and the projector works better at night. Compared with these time and space constraints, the mobile phone is more flexible and convenient because I can carry it with me anywhere and take it out to take a look at it at any time.''}. Another participant, P10, chimed in that head-mounted AR displays (HMDs) are generally expensive and have limited uses. He expressed his opinion by saying: \textit{``Although I was very impressed when I first tried Hololens 2, that feeling quickly faded away. I couldn't help but wonder if there was a good reason for me to spend over three thousand dollars on a new gadget that doesn't have much practical value. For me, the answer seems to be no.''}.
On the other hand, 4 participants who pointed out the deficiencies of HHD, i.e., the display scope of a projector is much larger than a mobile phone screen, which can hinder the recognition of generated text in AR apps. For example, the participant P8 claimed that the display content on mobile phone was difficult to view while moving by saying: \textit{``I realized that it was difficult for me to stay focused on what was displayed on my phone when I was moving around, especially small text content.''}. Therefore, tasks or scenes that rely heavily on text generation are not suitable for HHD devices.
8 participants highlighted that privacy and accessibility are important considerations, especially in multi-user collaboration and sharing scenarios. For instance, the participant P5 proposed that AIGC should be able to make some adjustments according to different situations by saying: \textit{``Content privacy issues need to be taken seriously. For different scenarios or different display modes, the displayed media content can be displayed in layers according to different permissions. In the past, this matter was usually participated by humans, but now it may be handed over to AI for automatic processing.''}. Potential solutions would be adopted for our AIGC+AR project include combining multiple display methods to surpass their limitations and providing hierarchical permissions for different users based on AI identification and authentication.

Additionally,as the main carrier of information, we noted that there is a significant difference in user experience between 2D and 3D content generation in consequence of the interview conversations. For image generation, many participants recognized that 3D virtual images can greatly increase user immersion and improve system usability. For example, the participant P9 expressed interest in trying on 3D virtual clothes by saying: \textit{``The idea of AR virtual trying on clothes is not that new, it is very interesting but also a bit troublesome because the clothes to try on always need to be manually configured by humans. Now generative AI provides new possibilities, and it may be very interesting to change clothes by speaking.''}. Nevertheless, a part of participants also showed rejection of 3D content, such as the P6 above-mentioned who has 3D motion vertigo by saying: \textit{``I don't think 3D objects in AR are necessary for me until I find a solution to my vertigo, 2D and 3D objects look fake anyway''}.
Text generation, meanwhile, is more sensitive to display size and requires sufficient space or dynamic display methods like scrolling or refresh. One of participants, P7, said such statement: \textit{``After trying your GenerativeAIR, I found that when the text generated by AI becomes too much, it is very difficult for me to read. Firstly, because of the limited size of some display devices such as head-mounted display or the small mobile screen. Secondly, since the generated content is not designed into a good layout and interaction, I have no way to adjust them.''}. Alternatively, converting text interaction to audio interaction is a practicable solution. For example, applying ChatGPT for dialog generation and display on smaller mobile phone screens may not be user-friendly.  For example, the P7 also supplemented: \textit{``Of course, we don't have to use our eyes to see. For text, it is also feasible to communicate only by voice for me.''}.

\section{Design Space}\label{design}
Via clustering and merging, we have condensed the multiple factors gathered from the focus group interviews into three overarching categories, namely ``user'', ``function'', and ``environment''. These categories are widely recognized as fundamental considerations in the design of interactive systems, as noted by previous studies \cite{villegas2018characterizing, wang2022constructing, henricksen2006developing}. Our objective is to explore the design space of AIGC+AR by investigating the relationships among these categories. Starting from user-centered thinking, our design space is structured into three aspects: what functions do the user need from the system (user-function), how the environment provides feedback to the user (user-environment) and what are the differences in the needs of different users (user-user).

\subsection{User-Function Design}
The AIGC+AR system is envisioned to offer diverse functions for the unique user experience.

\paragraph{Enabling personalized experience.} \textit{``AIGC can bring great flexibility and complement our design through its runtime content generation capability. (P5)''} Previously the artifacts in AR applications are created before the distribution and if there are items needed but not included in the package, it will be impossible to have them at runtime. AIGC opens up the opportunity to create high-quality content at runtime, which is particularly ideal for AR applications, because with AR, users are not isolated from their current context, where a level of personalization is attractive. 

\paragraph{Empowering the constrained.} \textit{``AR systems can observe my surroundings and deliver the information to me, and this has a great potential that they can help to observe the world for people who can't. (P9 P10)''} The advancements in technology should promote equality instead of only benefiting the advantaged. People with disability may be unable to access certain form of context while their AR devices can and use AIGC to transform the information into the accessible form in real-time. There are existing work e.g.\cite{chen2020unblind} on using AI to help the people with disability from certain perspectives, and leveraging AR+AIGC can largely extend such efforts.





\subsection{User-Environment Design}
The interactivity with the environment makes AR different from VR, and also provides distinct design space for AIGC+AR systems.

\paragraph{Bridging the virtual and real.}
\textit{``Sometimes AR did not give me a feeling of reality, because the content cannot fit in the environment I was surrounded by. I understand the developers can't exhaustively design a collection of artifacts for every scenario, but it still somehow ruined the experience (P1 P7)''} The virtual scene and real environment are separated by a gap that breaks the fidelity and immersion of user experience. This gap can be filled with AIGC. One way is by assisting the content generation, through adapting the initial designed artifacts to fit the environment ("AI Assist" in \ref{RQ2}); and another way is by generating new elements for the unprepared environment ("AI Generation" in \ref{RQ2}). Dynamically combining
these two methods can enable a seamless experience through connecting the virtual scenes and real environments.

\paragraph{Accommodating dynamic factors.}
There are multiple dynamic factors involved in the AR system, typically the different AR display methods, the distinct scenarios, and various environmental factors. These can be catered by integrating AIGC into the system. For such a system, the feedback supplied by the environment to the user is mainly dependent on the presentation of the AR display. As mentioned above, the three AR display methods (SAR, HMD and HHD) have separate display performances, particularly in terms of functionality, portability and privacy. \textit{``The potential application scenarios of HHD based on mobile phones are vast and varied. (P3)''}. Alternatively, in different scenarios (e.g., indoors, outdoors, working and entertainment), environmental factors (e.g., light and sound) should be included in the design as well, which is beneficial for ameliorating user experience such as immersion and interactivity. AIGC can be leveraged to address such requirements. \textit{``It would be fun if a machine could somehow know my current mood and adapt the content and style of the generated image accordingly. (P4)''}.

\subsection{User-User Design}
AR systems should be aware of other users within the same space, and AIGC can serve as a piece of this puzzle.

\paragraph{Building shared experience.}

One angle of user-user design is how AR+AIGC systems can coordinate to generate content for better content sharing and collaboration. The shared space can be complex and leveraging AIGC can avoid users being isolated. Specifically, users may have different roles in the same shared space, and have different needs. For instance, in Hartmann's work \cite{hartmann2020aar}, the concepts of ``presentation user'' and ``external user'' were introduced. These two types of users engage with media content asymmetrically. For ``presentation user'', who conducts the AR device for presenting purpose, they care more about portability and privacy. For ``external users'', their main needs are immersion, low cognitive cost and communication with other users. Therefore, both of AI part and AR part need to be coordinated for distinctive needs for multiple users in the same shared space, such as hierarchical user accessibility based on AI recognition and authentication, or content sharing and switching based on different AR display methods. \textit{``If the generated content involves my privacy, such as my photo albums and life vlog, I would like AI to understand which content can be shown to my friends. If it generates or shows what I don't want others to see content, it would be too embarrassing. (P2)''}.

\section{Potential Applications}\label{application}

\begin{figure}
\centering
\includegraphics[width=0.5\textwidth]{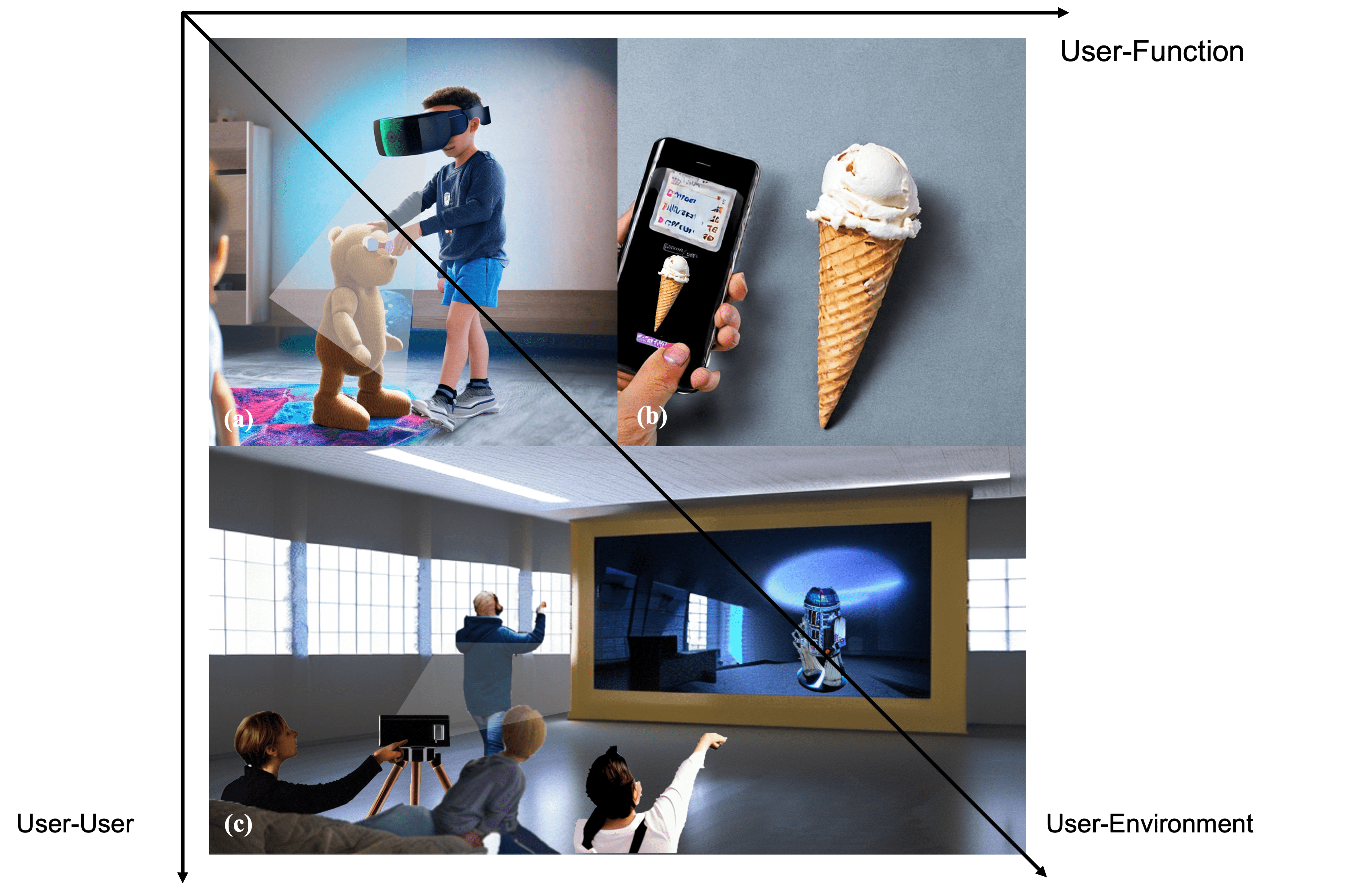}
  \caption{Potential Applications of GenerativeAIR: (a) Boosting real-time creative media generation; (b) Smoothening interactions with surroundings and environment; (c) Facilitating multi-user collaboration. }
  ~\label{fig:application}
\end{figure}


In this section, we summarize and highlight three streams of potential applications enabled by AR+AIGC. Firstly, generative AI models can enhance real-time creative media generation with personalization, as illustrated in Figure \ref{fig:application} (a), where a boy wearing AR glasses interacts with a virtual teddy bear. Such generation can be largely personalized from various perspectives to cater the needs of users. One example is AR fitting room, which can be significantly improved functionally from the integration of AIGC. Moreover, people with disability, the minority, or people from unprivileged groups can also benefit from relevant applications with enhanced capabilities from the integration of AIGC. Secondly, AR+AIGC unlocks the potential of more smooth interactions with the environment and surroundings, as shown in Figure \ref{fig:application} (b). For example, AR games may become more realistic and AR pets may become more lively. Furthermore, GenerativeAIR can enable a better shared experience, and address privacy and privilege classification issues in multi-user scenarios by assigning hierarchical display content based on user permissions and privacy levels through id-authentication, as depicted in Figure \ref{fig:application} (c).

To interpret these application scenarios, we employ a user-centered design approach as discussed in section \ref{design}. Specifically, (a) and (b) are intended for single users, while (c) is designed for multiple users. GenerativeAIR offers various functions to meet diverse requirements, and the interaction between user and environment remains a persistent theme.

\section{Limitations and Future Work}\label{future_work}

This work aims to explore the potential design space of generative AI for using a simple prototype GenerativeAIR. The limitations of our work primarily lie in the technical aspects that require further improvement. For example, our current prototype only generates 2D images using the Stable Diffusion model, and we acknowledge the potential benefits of generating 3D content in AR. Additionally, our prototype is currently offline and lacks real-time interaction, hindering its practical application. Future work will focus on implementing real-time functionality and integrating additional software and hardware to enrich the system's functions. Moreover, we have not addressed the hierarchical difficulty of privacy and permissions in multi-user scenarios, which is a critical issue for collaborative and sharing settings.





\section{Conclusion}
This paper introduces the concept of AIGC+AR and explores its design space. We first construct a prototype by integrating two text-input generative AI models with three common AR displays in section \ref{prototype}. More details about our focus group could be seen in section \ref{methodology}. Next, in section \ref{discussion}, we provide a qualitative comparison of the advantages and disadvantages of our work compared to related studies. Also we discuss factors that need to be considered in technology development itself. Furthermore, a ``user-fucntion-environment'' design thinking is proposed and discussed section \ref{design}. Last, in section \ref{application}, we present and analyze potential application scenarios.

\bibliographystyle{ACM-Reference-Format}
\bibliography{sample}


\end{document}